\documentclass{elsart}
\usepackage{graphicx}
\begin{document}

\begin{frontmatter}
  \title{
    Development of a Large-Area Aerogel \v{C}erenkov Counter 
    Onboard BESS
    }
  \author[Tokyo]{Y.~Asaoka},
  \author[Tokyo]{K.~Abe},
  \author[Tokyo]{K.~Yoshimura},
  \author[Kyoto]{M.~Ishino},
  \author[Tokyo]{M.~Fujikawa},
  \author[Tokyo]{S.~Orito}
%  \author[Tokyo]{Youichi Asaoka},
%  \author[Tokyo]{Koh Abe},
%  \author[Tokyo]{Koji Yoshimura},
%  \author[Kyoto]{Masaya Ishino},
%  \author[Tokyo]{Motoharu Fujikawa},
%  \author[Tokyo]{Shuji Orito}
  \address[Tokyo]{
    Department of Physics, School of Science, University of Tokyo, Japan
    }
  \address[Kyoto]{
    Department of Physics, School of Science, University of Kyoto, Japan
    }
  \begin{abstract}
    This paper describes the development of a threshold type aerogel
    \v{C}erenkov counter with a large sensitive area of $0.6 \, {\rm
    m^2}$ to be carried onboard the BESS rigidity spectrometer to
    detect cosmic-ray antiprotons. 
    The design incorporates a large diffusion box containing 46
    finemesh photomultipliers, with special attention being paid
    to achieving good performance under a magnetic field and providing
    sufficient endurance while minimizing material usage. 
    The refractive index of the aerogel was chosen to be $1.03$.
    By utilizing the muons and protons accumulated during the
    cosmic-ray measurements at sea level, a rejection factor of
    $10^4$ was obtained against muons with $\beta \approx 1$, while
    keeping 97\% efficiency for protons below the threshold.
  \end{abstract}
\end{frontmatter}

\section{Introduction}
In three consecutive scientific flights over northern Canada
(1993--1995), the balloon-borne BESS rigidity spectrometer, which has a
large acceptance of $0.3 \,{\rm m}^2{\rm sr}$, precisely measured the
rigidity, velocity, and d$E\!/$d$x$ of cosmic-ray antiprotons.
The first definite (mass-identified) detection of low-energy
antiprotons was achieved~\cite{kn:pbar93} by using BESS'93 data in the
kinetic energy range from $0.3$ to $0.6 \,{\rm GeV}$.
By improving the time-of-flight (TOF)  resolution, the BESS'95 data
further led to background-free detection of $43$ antiprotons in the
range from $0.18$ to $1.4 \,{\rm GeV}$~\cite{kn:pbar95}.
The resultant energy spectrum appeared flat below $1 \,{\rm GeV}$,
where the production of secondary antiprotons in interstellar space is
expected to sharply decline.
Although such phenomena could very well indicate the existence of an
admixture of primary antiprotons from novel
sources~\cite{kn:pbh,kn:pbh2} such as the evaporating
primordial black holes or the annihilation of neutralino dark
matter, present statistical accuracy and the limited energy range do
not allow us to draw a firm conclusion due mainly to the uncertainties
in the calculation of secondary antiprotons. 

As a means of confirming whether or not a primary antiproton signal
does exist, the absolute flux of secondary antiprotons must be determined by
expanding the detection range up to about $3 \,{\rm GeV}$
kinetic energy; a range covering the expected peak (at $\sim 2.5
\,{\rm GeV}$) in the secondary antiproton flux, which is generally
considered to be the predominant source of antiprotons.

Independent to the search for the novel primary antiproton component,
the precise measurement of the secondary antiprotons itself will be of 
crucial importance to determine the propagation mechanism of cosmic
rays in the Galaxy.

Based on above scientific considerations, one of us proposed~\cite{kn:prj}
achieving background-free and positive (mass-identified) detection of
antiprotons up to $3 \,{\rm GeV}$ by ({\em i}) eliminating the
overwhelming $e^- /\mu^- $ background with a threshold-type aerogel
\v{C}erenkov counter having proper refractive indices around 1.03, and 
({\em ii}) improving the timing resolution of each TOF counter to $50
\,{\rm ps}$.

In this paper, the design, construction, and performance of the aerogel 
\v{C}erenkov counter are presented.

\section{General description}\label{sec:acc}
\subsection{Configuration onboard BESS '97 spectrometer}
\begin{figure}[hbtp]
  \begin{center}
    \includegraphics[width=12cm]{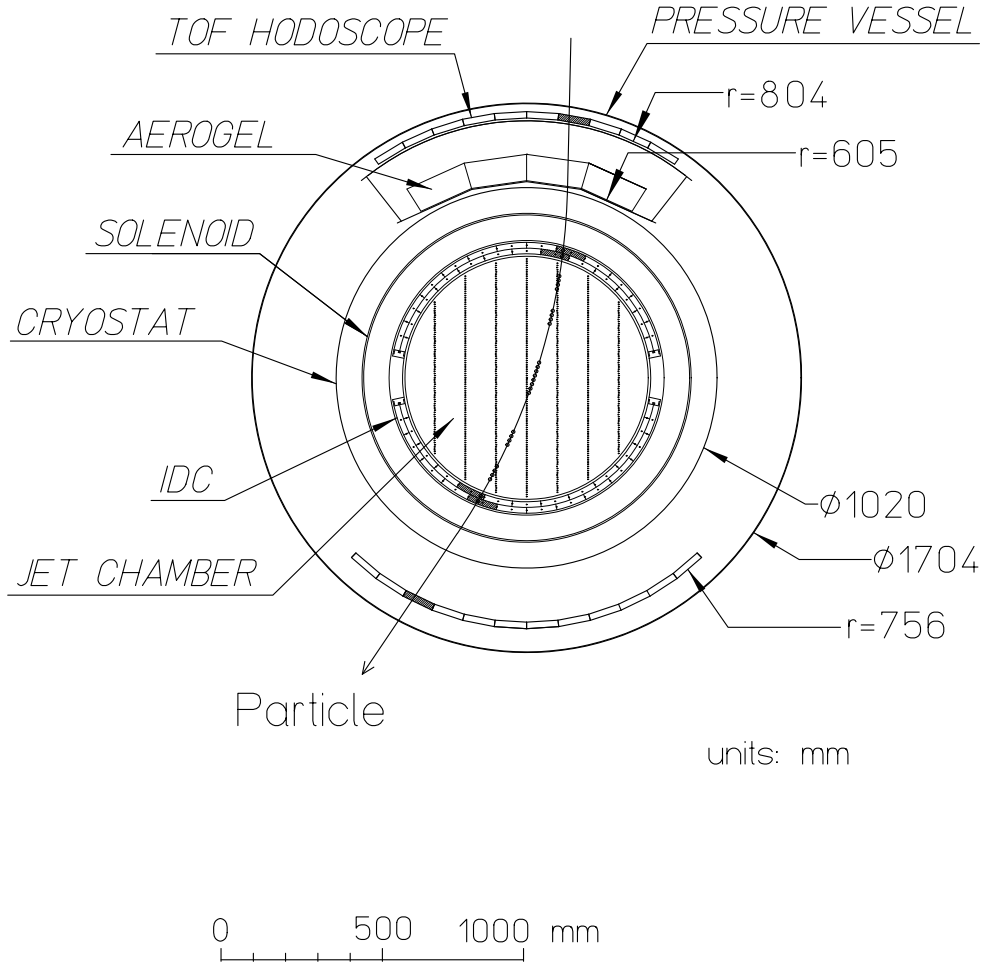}
  \end{center}
  \caption
  [Cross-sectional diagram of the BESS'97 spectrometer showing
  a negatively-charged particle event.]
  {Cross-sectional diagram of the BESS'97 spectrometer showing
  a negatively-charged particle event.}
\label{fig:bess}
\end{figure}

Figure~\ref{fig:bess} shows a cross-sectional diagram of '97 version
of the BESS rigidity spectrometer (for detail, see~\cite{kn:pbar93,kn:det})
consisting of a superconducting solenoidal magnet with a central
nominal field of 1 T, inner drift chamber (IDC), JET drift chamber,
TOF plastic scintillation hodoscope, and aerogel \v{C}erenkov counter. 
Onboard accommodation of the aerogel \v{C}erenkov counter was achieved 
by positioning it between the upper TOF hodoscope and cryostat.

\subsection{BESS aerogel \v{C}erenkov counter}
In describing the developed counter, cylindrical coordinates $( r,
\phi, z )$ are used in which the magnetic field direction is taken as
the {\em z}-axis and the perpendicular plane as the $r\phi$ plane.
The following design  constraints were applied to the counter.
\begin{enumerate}
  \item The counter should have a large solid angle and area such that 
    it matches the large acceptance of the BESS
    spectrometer. \label{itm:lrg}
  \item The counter should have a thickness of $\leq 19 \,{\rm
      cm}$. \label{itm:thc}
  \item The counter should operate in the fringe magnetic field of
    $\sim 0.2 \, {\rm T}$. \label{itm:mag}
  \item To determine the threshold of antiprotons to about $3 \,{\rm
      GeV}$, the aerogel refractive index ($n$) should be $\leq
      1.035$. \label{itm:ind}
  \item The rejection factor between antiproton and background 
    $e^-/\mu^-$ should be at least $10^3$ in order to have 
    background-free detection of antiprotons. \label{itm:rej}
  \item The integrity of aerogel blocks should be maintained against
    shock and vibration during shipping, launching, and parachute
    landing. \label{itm:itg}
\end{enumerate}

To verify that design constraints ~(\ref{itm:lrg})--~(\ref{itm:itg})
could be satisfied, we built a prototype counter $( 22 \times 96
\times 21 \,{\rm cm}^3 )$ incorporating 14 finemesh photomultiplier
tubes (PMTs) at each end, being about one-fourth in size relative to
the envisioned counter.
Measurements of cosmic-ray muons were taken such that various
configurations could be examined to optimize reflector material, the
placement of slanted end plates on which the PMTs would be mounted, and
aerogel thickness.

\begin{figure}[hbtp]
  \begin{center}
    \includegraphics[width=12cm]{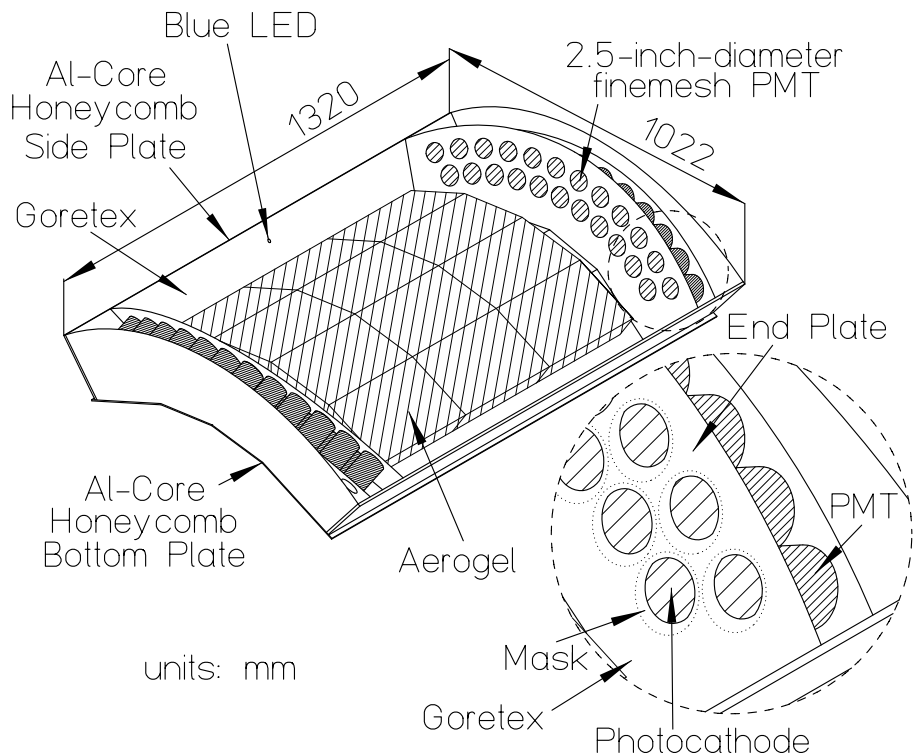}
  \end{center}
  \caption
  [Overview of the aerogel \v{C}erenkov counter.]
  {Overview of the aerogel \v{C}erenkov counter.}
  \label{fig:acc}
\end{figure}

Figure~\ref{fig:acc} shows an overview diagram of the final design,
where the counter consists of a large diffusion box containing aerogel
blocks viewed by $46$ PMTs densely arranged at the both ends.
The unit's weight and the amount of material were minimized using an
aluminum-core honeycomb plate as the main structure.

Choosing an effective reflector material is a key aspect in counter
design since most photons generated in the diffusion box will undergo
numerous reflections prior to reaching the PMTs.
We accordingly tested various type of sheet material, {\em i.e.},
Millipore, Tibex, and Goretex; ultimately finding that Goretex, which
exhibits high reflectivity in the short-wavelength region ($300$--$400
\,{\rm nm}$), is most suitable.
From the standpoint of photon collection, this is consistent with the
fact that the number of \v{C}erenkov photons is inversely
proportional to the square of wavelength; thereby making reflectivity
in the short-wavelength region important.

In consideration of operating the counter in a $0.2$-${\rm T}$ fringe
magnetic field, we selected $2.5$-inch-diameter finemesh type PMTs
(R6504S, Hamamatsu Photonics K. K.) whose sensitive region lies
between $300$ and $700 \,{\rm nm}$.\@
Readout electronics consists of summing amplifiers that combines 46
PMT signals into 8 channels, which are digitized by a
charge-integrated ADC.\@
Blue LEDs (NLPB, NICHIA) with a peak of $450 \,{\rm nm}$ are used to
adjust the PMT gain such that all the PMTs provide the same ADC counts
per photoelectron.
Since PMT gain shows magnetic field dependence, final high-voltage
tuning must be done in the counter after exciting the solenoidal
magnet to the nominal field.
Therefore, the blue LEDs were mounted on the both sides of the side plate
at its center point; a configuration allowing PMT gain to be monitored
throughout the experiment. 

Another consideration concerns the magnetic field itself, since
if the PMT axis and magnetic field direction form a nonzero angle $\theta$, 
then the PMTs lose their effective photocathode area ($S_{\rm eff}$),
{\em i.e.}, some secondary electrons produced at dynode, traveling
inside the PMT in the direction of the magnetic field, cannot reach
the anode.
Accordingly, to avoid losing $S_{\rm eff}$, each end plate on which
PMTs were mounted was slanted to reduce $\theta$.\@ 
The drawback in this approach, however, is that slanting in turn
leads the loss of photoelectrons.
It is for this reason that during prototype testing we focused our
attention on optimizing the angle of slanting such that photoelectron
loss is at an acceptable level.
The slanting angle was determined to be $24.5^{\circ}$ after
optimizing photoelectron loss and $S_{\rm eff}$.  
This slanting angle reduced $\theta$ from
$39^{\circ}$ to about $18^{\circ}$ in the lower PMTs and from
$43^{\circ}$ to about $26^{\circ}$ in the upper PMTs, while $S_{\rm eff}$
increased by 20\% in comparison with not slanting.
A mask of Goretex was placed over the front surface of each PMT
(Fig.~\ref{fig:acc}) such that the photons hitting the insensitive
photocathode area formed by residual $\theta$ would be reflected back.

\subsection{Aerogel}\label{acmnt}
As a \v{C}erenkov radiator, we selected silica aerogel (Mori-Seiyu
Co.) having a refractive index of 1.032, which was optically measured
in several pieces of sample using a He-Ne laser with wavelength of
632.8 nm~\cite{kn:bagel}.
This aerogel was manufactured using a new method~\cite{kn:bagel} which 
ensures that it retains its hydrophobicity such that long-term
stability and good clarity are afforded.
It is the aerogel's excellent clarity that allowed us to use a
diffusion {\em versus} mirror box design despite being constrained by
a counter thickness of $\leq 19 \,{\rm cm}$.\@

\begin{figure}[hbtp]
  \begin{center}
    \includegraphics[width=10cm]{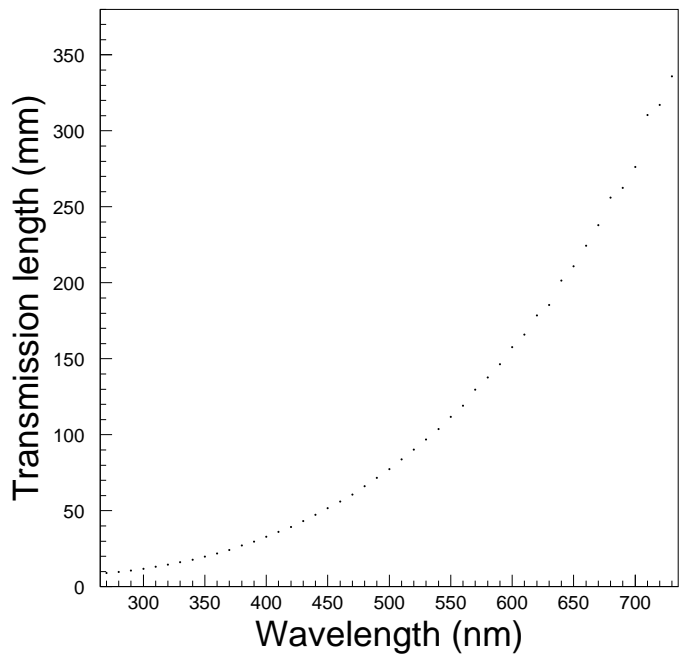}
  \end{center}
  \caption
  [Transmission length as a function of the wavelength obtained with a 
  photospectrometer.]
  {Transmission length as a function of the wavelength obtained with a 
  photospectrometer.}
  \label{fig:trans}
\end{figure}

Transmission of the aerogel samples were measured using a
photospectrometer~\cite{kn:bagel}. 
The obtained data was translated into the transmission length
($\Lambda$) by the function:
\[ T = T_0 \exp( -d/\Lambda ) \, , \]
where $T$, and $d$ are the transmission and thickness of aerogel,
respectively.
Figure~\ref{fig:trans} shows a typical sample of wavelength dependence
of the transmission length for a 2.1-cm thick sample.
The transmission length strongly depends on the wavelength.

The aerogel's unique hydrophobicity enabled it to be manufactured by a
water-jet cutter to the correct size with an accuracy of $0.5 \, {\rm
  mm}$.
Each block of aerogel was cut with a slight taper along its width
(toward the $\phi$ direction) in order to minimize gaps after
assembly.

Aerogel can be configured by either suspending blocks of it from 
the top plate or attaching blocks to the bottom plate.
We had surmised that the former configuration would lead to detecting
a higher number of photoelectrons, although aerogel fixation would be 
relatively more difficult.
However, prototype testing showed no significant difference in the
measured number of photoelectrons for the two cases; thus we attached
aerogel blocks to the bottom plate.
PMTs were mounted on the upper part of each end plate since in
prototype testing this configuration detected about 10\% more
photoelectrons {\em versus} a configuration in which PMTs directly
viewed the aerogel blocks.

Onboard space limitations dictated the design constraint of a
$19$-cm-thick counter. 
The prototype tests showed that when the thickness of aerogel reached
half that of the counter, the measured number of photoelectrons was at
saturation; whereas at three-fourths thickness, the number decreased
due to aerogel absorption of photons. 
With these results in mind, and since our main target is low-energy
antiprotons, the aerogel thickness should necessarily be as thin as
possible in terms of quantity of material used.
At a density of $0.128 \,{\rm g}/{\rm cm}^3$, the thickness of the aerogel
blocks was determined to be $ 8 \,{\rm cm}$.\@

\begin{figure}[hbtp]
  \begin{center}
    \includegraphics[width=10cm]{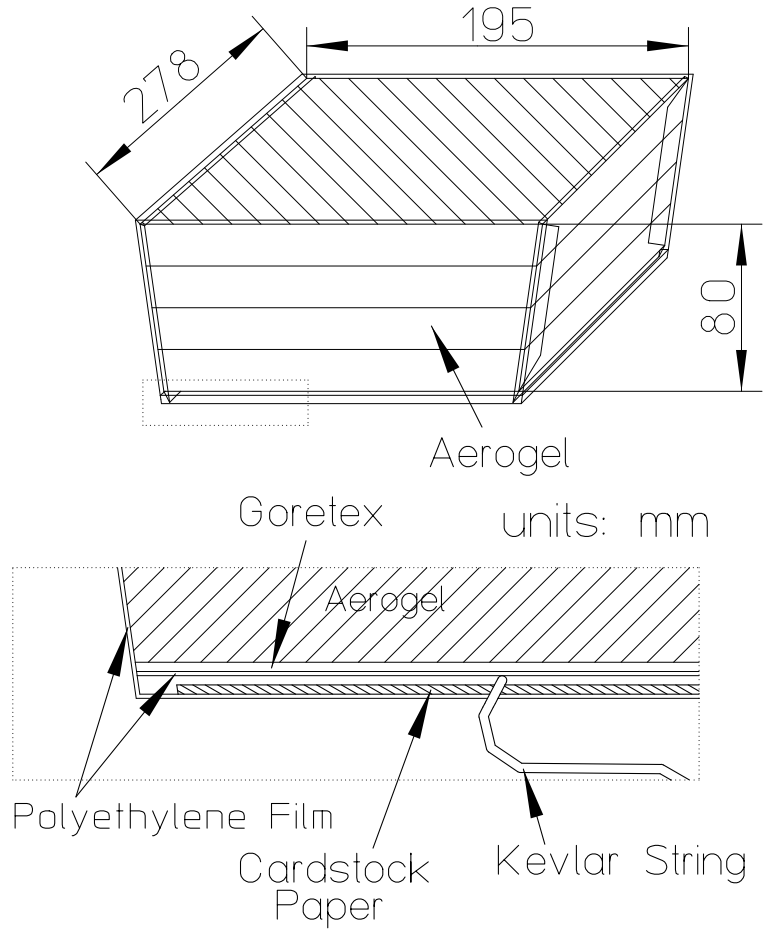}
  \end{center}
  \caption
  [Overview of an aerogel block.]
  {Overview of an aerogel block.}  
  \label{fig:wrap}
\end{figure}

Figure~\ref{fig:wrap} shows how an aerogel block was made,
{\em i.e.}, a stack of four $2$-cm-thick layer pieces with a piece of
Goretex sheet on their bottom were wrapped lengthwise with
polyethylene film (ITOCHU SANPLUS Co. Ltd.).\@
Then, after placing cardstock paper on the bottom side, it was again
wrapped around its width.
Polyethylene film was chosen as it has good clarity in the relevant
wavelength region.
A string made of Kevlar was used to fix (tie) the aerogel block to the
bottom honeycomb plate through a hole drilled on it. 
With this configuration of block fixation, no photon-absorbing
surfaces such as aluminum support surfaces are present within the
diffusion box. 

When assembled, the aerogel blocks form upper and lower surfaces of
$0.65$ and $0.57 \, {\rm m^2}$, respectively.

\section{Performance}\label{sec:anly}
\begin{figure}[hbtp]
  \begin{center}
    \includegraphics[width=10cm]{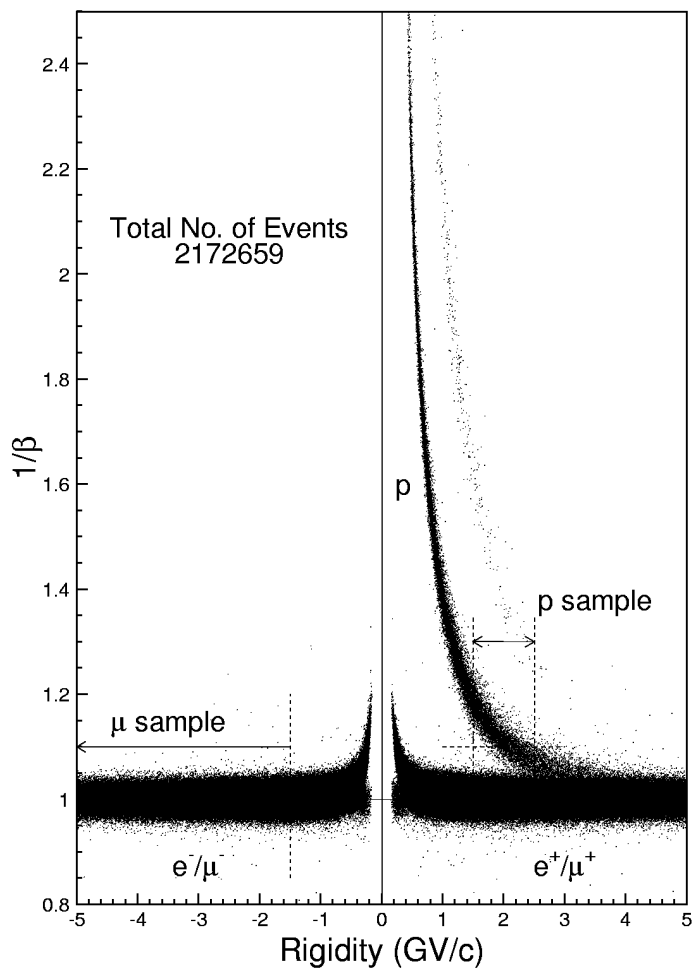}
  \end{center}
  \caption
  [$1/\beta$ vs $Rigidity$]
  {Scatter plot of $1/\beta$ vs rigidity using cosmic-ray data at sea
  level.} 
  \label{fig:bvsr}
\end{figure}

Counter performance was evaluated using cosmic-ray data collected
with the full BESS'97 configuration at sea level over a 4-d period in
May 1997. 
Figure~\ref{fig:bvsr} shows the obtained scatter plot of $1/\beta$
{\em versus} spectrometer rigidity, where proton ($p$) and muon ($\mu$)
bands are clearly separated up to $2.5 \, {\rm GV}\! /c$, being
identified by rigidity and TOF measurements (mass determined).\@

\subsection{Counter response}\label{sec:uni}
From sea level data of negatively charged muons in the
rigidity range of $-10$ and $-2 \,{\rm GV}\! /c$, we evaluated the
response of the counter at $\beta \approx 1$.

\begin{figure}[hbtp]
  \begin{center}
    \includegraphics[width=12cm]{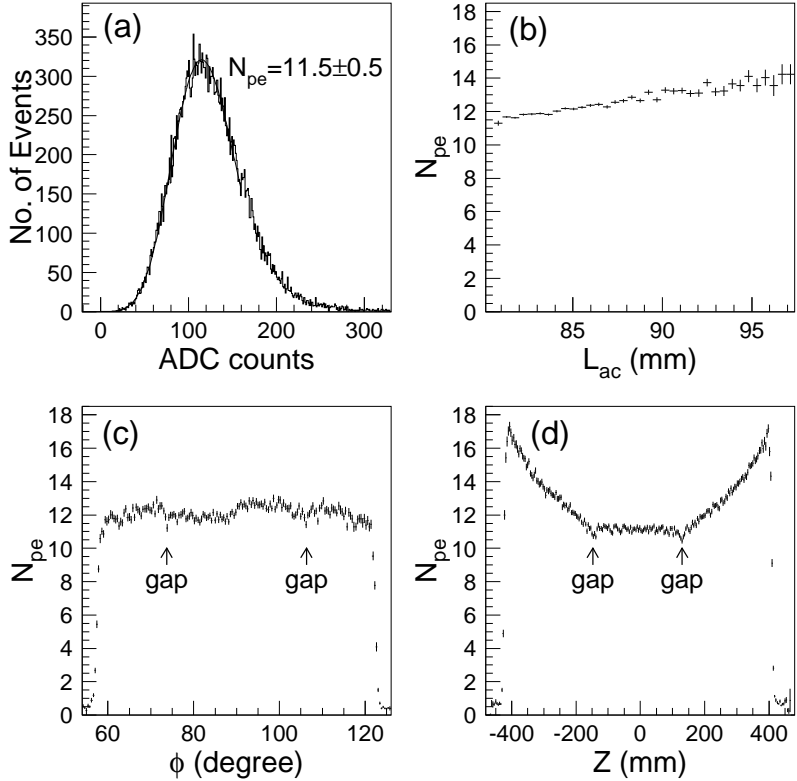}
  \end{center}
  \caption
  [Uniformity study]
  {(a) ADC distribution around the center of the counter with a
  superimposed fit performed between 0 and 260 ADC counts. 
  (b)--(d) respectively represent $L_{\rm ac}$, $\phi$, and $z$
  dependence of $N_{\rm pe}$.}
  \label{fig:dep}
\end{figure}

Figure~\ref{fig:dep}(a) shows the charge (ADC) distribution
(histogram) obtained around the center of the counter for vertically
incident muons.
By fitting data in region 0 to 260 counts to an
approximation~\cite{kn:cpois} of a Poisson distribution, the
effective~\cite{kn:sim,kn:fmpm} mean number of photoelectrons ($N_{\rm pe}$)
was determined to be $11.5 \pm 0.5$.\@

The position and angle dependence of the incident particle was
measured over the counter's sensitive region.
Figure~\ref{fig:dep}(b)--(d) shows $N_{\rm pe}$ {\em versus} path length in
aerogel ($L_{\rm ac}$), $\phi$, and $z$, respectively.
These figures indicate that $N_{\rm pe}$ is proportional to $L_{\rm
  ac}$, and that $N_{\rm pe}$ is dependent on $z$ of the incident
particle whereas $\phi$ dependence of the counter is small.
Note that only slight decreases of $N_{\rm pe}$ are present at the
positions corresponding to the gaps between assembled aerogel blocks.
The influence of these gaps can be totally eliminated during the
offline analysis stage using geometrical cutting 
while maintaining the cut efficiency at more than 99\%.\@  

\begin{figure}[hbtp]
  \begin{center}
    \includegraphics[width=10cm]{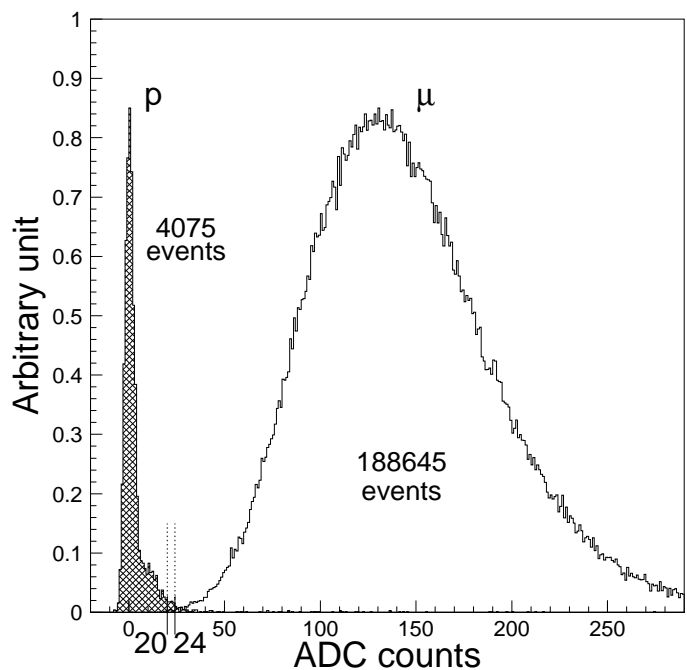}
  \end{center}
  \caption
  [Summed ADC spectrum]
  {ADC distribution obtained by summing all PMT signals, where results 
    are shown corresponding to muons in the rigidity range from $-10$ to
  $-2 \,{\rm GV}\! /c$, and protons from $1.5$ to $2.5 \,{\rm GV}\! /c$.} 
  \label{fig:hac}
\end{figure}

In order to select the particles which passed through the aerogel, we
applied the fiducial cut by requesting that the extrapolated track
should cross the upper as well as lower surface of the aerogel.
The open histogram in Fig.~\ref{fig:hac} shows the counter's response
throughout its sensitive region to muons at $\beta \approx 1$,
which allows the rejection factor of the antiproton background to be
estimated. 
We also studied the response of the counter to below-threshold particles.
The hatched histogram in Figure~\ref{fig:hac} shows the charge
distribution for protons in the rigidity range from $1.5$ to $2.5 \,
{\rm GV}\! /c$.
To avoid muon contamination, only $ 1/\beta > 1.1 $ particles were
used (Fig.~\ref{fig:bvsr}).\@
From a detailed analysis, the upper tail of the histogram produced by
below-threshold particles can be attributed to several different sources:
\begin{enumerate}
\item \v{C}erenkov light produced in the Goretex and polyethylene
  wrapping; \label{itm:che}
\item scintillation light produced in aerogel, Goretex, and
  polyethylene wrapping; \label{itm:sci}
\item $\delta$-rays; \label{itm:del}
\item interactions in the BESS spectrometer; \label{itm:int}
\item contribution by accidental particles. \label{itm:aci}
\end{enumerate}
The signal produced by sources ~(\ref{itm:che})--~(\ref{itm:del}) was
investigated during prototype testing and found to be negligibly small
($N_{\rm pe}<1$) or rare enough to keep
identification efficiency for protons high. 
Contamination by sources ~(\ref{itm:int}) and ~(\ref{itm:aci}) can
be eliminated in offline analysis by the use of track quality cuts.

\subsection{Estimation of efficiency and contamination levels}
\begin{figure}[hbtp]
  \begin{center}
    \includegraphics[width=10cm]{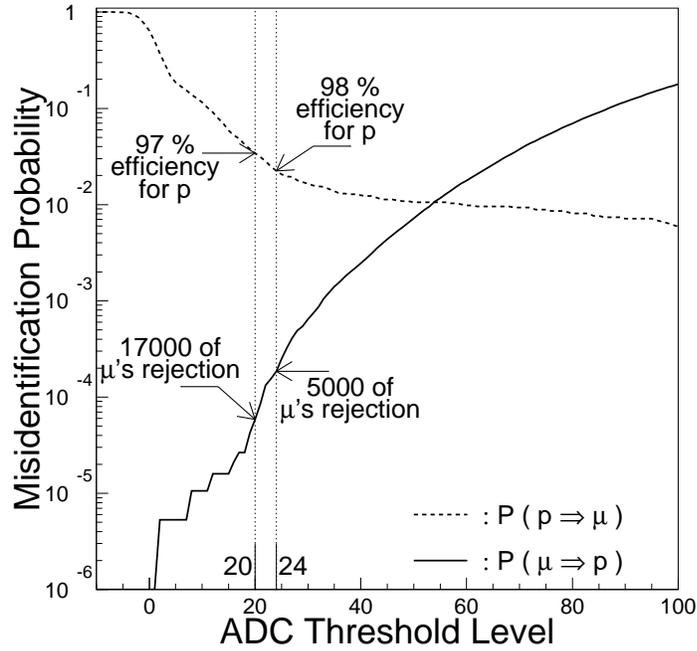}
  \end{center}
  \caption
  [Rejection and Efficiency]
  {Misidentification probability $\wp(p \Rightarrow \mu)$ and 
    $\wp( \mu \Rightarrow p)$ as a function of charge-threshold
    levels.}
  \label{fig:eff}
\end{figure}

To estimate counter performance in terms of separation power, the
results of muon and proton detection in Fig.~\ref{fig:hac} were
combined such that the following quantities could be obtained as a
function of the ADC threshold level:  
\begin{itemize}
\item  the misidentification probability $\wp(p \Rightarrow \mu)$ that
  a proton gives a higher signal than the threshold; \label{itm:p2m}
\item  the misidentification probability $\wp( \mu \Rightarrow p)$
  that a muon gives a lower signal than the threshold. \label{itm:m2p}
\end{itemize}
From Fig.~\ref{fig:eff} summarizing the results, if the threshold is
set at 20 as shown, then a $1.7 \times 10^4$ rejection factor of
muons is achieved while keeping the efficiency for protons at 97\%.\@

\subsection{Estimation of aerogel index}
\begin{figure}[hbtp]
  \begin{center}
    \includegraphics[width=10cm]{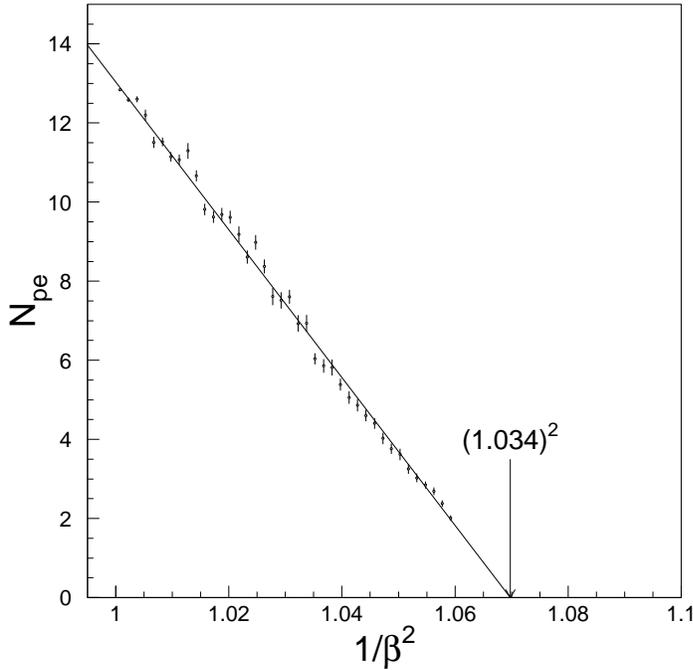}
  \end{center}
  \caption
  [Muon threshold curve]
  {Momentum dependence of $N_{\rm pe}$ as a function of 
    $1/\beta^{2}$.}
    \label{fig:index}
\end{figure}

To determine the momentum threshold of the aerogel \v{C}erenkov
counter, the momentum dependence of $N_{\rm pe}$ was measured using 
muons identified by mass determination.
Figure~\ref{fig:index} shows a plot of  $N_{\rm pe}$ {\em versus}
$1/\beta^{2}$ calculated using
  \[\frac{1}{\beta^2} = \frac{m^2}{R^2}+1 \, ,\]
where $R$ is the rigidity of the incident particle and $m$ is the
mass of a muon.
$1/\beta^{2}$ was calculated using rigidity vice TOF measurements
because the resultant value is more accurate in cases where the mass
of the incident particle is known.
The superimposed line represents a linear fit to the data.
At $N_{\rm pe} = 0$, this corresponds to the square of the
refractive index, {\em i.e.}, $n=1.034\pm0.001$, which is
slightly larger than the optically measured value of 1.032.
This discrepancy in $n$ might be due to the effect of below-threshold
particles (Sec.~\ref{sec:uni}).\@

\section{Summary and conclusions}\label{sec:sum}
We have described the development of an aerogel \v{C}erenkov counter
designed for use onboard BESS. This counter features a large
sensitive area, sufficient flight durability, and good performance
in a magnetic field of 0.2 T.\@
Sea level tests in which muons and protons were detected showed that a
rejection factor of $10^4$ is achieved while maintaining
identification efficiency at 97\% for below-threshold particles. In
terms of the threshold of the employed aerogel, it is estimated that
antiprotons can be detected up to $3.6 \, {\rm GV}\! /c$.

\begin{ack}
  S.~O. sincerely thanks to H.~Sato and K.~Kurosawa for contributions early
  in the R/D stage, being crucially important for 
  the initialization of the work described in this paper.
  We appreciate supports in the manufacturing processes of the aerogel
  by Y.~Shikaze, T.~Yoshida, and Y.~Watanabe, and valuable help by
  M.~Motoki, H.~Matsunaga, T.~Sanuki, M.~Sasaki, and other members of
  the BESS collaboration.
  This work was supported by a Grant-in-Aid for Scientific Research 
  from the Japanese Ministry of Education, Science and Culture.
  The analysis was performed using the computing facilities
  at ICEPP, Univ. of Tokyo.
\end{ack}


\begin{thebibliography}{99}
  \bibitem{kn:pbar93} K.~Yoshimura et al.,
    Phys. Rev. Lett. 75 (1995) 3792; 
    A.~Moiseev et al., Astrophys. J. 474 (1997) 479.
  \bibitem{kn:pbar95} H.~Matsunaga et al., in preparation.
  \bibitem{kn:pbh} K.~Maki, T.~Mitsui, S.~Orito,
    Phys. Rev. Lett. 76 (1996) 3474;
    T.~Mitsui, K.~Maki, S.~Orito, Phys. Lett. B 389 (1996) 169.
  \bibitem{kn:pbh2} S.~Orito, T.~Mitsui, K.~Maki, in preparation.
  \bibitem{kn:prj} S.~Orito, KEK Proceedings 96-9 (1996) 105, edited by
  A.~Yamamoto and T.~Yoshida.
  \bibitem{kn:det} K.~Anraku et al., in preparation.
  \bibitem{kn:bagel} I.~Adachi et al., Nucl. Instr. and Meth. A 355
    (1995) 390.
  \bibitem{kn:cpois} L.~C.~Alexa et al., Nucl. Instr. and Meth. A 365
    (1995) 304.
  \bibitem{kn:sim} R.~Suda et al., Nucl. Instr. and Meth. A 406 (1998) 
    213.
  \bibitem{kn:fmpm} R.~Enomoto et al., Nucl. Instr. and Meth. A 332
    (1993) 129.
\end{thebibliography}
\end{document}